**Title:** Probabilistic linkage with graphs: leveraging cluster structure to improve the accuracy of entity resolution at scale

**Authors:**

Evelyn Lauren[1,2], Laura White[2], Trevor Graham Bell[3], Jacob Bor[1,4,5,6]

[1]Health Economics and Epidemiology Research Office, Faculty of Health Sciences, University of the Witwatersrand, Johannesburg, South Africa
[2]Department of Biostatistics, Boston University School of Public Health, Boston, MA, USA
[3]National Institute for Communicable Diseases, National Health Laboratory Service, Johannesburg, South Africa
[4]Africa Health Research Institute, Somkhele, South Africa
[5]Department of Global Health, Boston University School of Public Health, Boston, MA, USA
[6]Department of Epidemiology, Boston University School of Public Health, Boston, MA, USA

**Abstract**

Probabilistic record linkage identifies records belonging to the same entity even when identifiers are incomplete or imperfect, and is increasingly used to conduct population-level inferences from administrative data at scale. Linkage produces noisy pairwise links that must then be resolved into distinct individuals. The standard approach, transitive closure, applies a fixed threshold to each link independently and assigns clusters from the resulting connected components. However, the risk of merging distinct individuals grows with population size, motivating approaches that use graph structure to improve linkage accuracy. Clustering algorithms from adjacent fields solve the same graph-clustering task in other domains, yet have never been systematically evaluated for entity resolution. We evaluated eleven clustering algorithms – five established in entity resolution and six from adjacent fields – across three simulated datasets and one real-world dataset, and propose an adaptation tailored to linkage graphs. Cross-field algorithms – Leiden, Louvain, label propagation – were among the most accurate, reducing linkage error by up to 66% relative to transitive closure. The three algorithms also ran up to 40 times faster than Markov clustering (a leading approach in the entity resolution literature) and integrate directly into existing pipelines at scale.

## INTRODUCTION

### Graph-based entity resolution

Administrative data are increasingly used in research as they enable large-N population-level inferences at relatively low cost[1,2]. However, administrative data may contain incomplete or imperfect identifying information (e.g., misspellings), making it difficult to determine which records correspond to the same underlying individual. Probabilistic record linkage addresses this by computing similarity scores for record pairs based on agreement across identifying attributes such as names and demographic information[3]. These scores are then used to partition records into clusters each representing a unique individual, a process known as entity resolution[4,5].

In deduplication settings, where multiple records (within or across datasets) correspond to a single individual, the resulting similarity structure forms a network (i.e. graph) in which nodes represent records and weighted edges represent pairwise similarity scores. Records belonging to the same entity tend to form dense clusters with high-scoring edges, with only sparse, low-scoring links connecting records from different entities. The goal of entity resolution is to recover these underlying clusters such that each cluster corresponds to a unique individual. The traditional approach (implemented in popular data linkage packages such as Splink[6]) forms clusters as connected components based on the property of transitive closure. All record pairs with similarity scores above a fixed threshold are classified as matches; if A links to B and B links to C, then A, B, and C are all in the same cluster. While simple and computationally efficient, this approach ignores potentially valuable information encoded in the network structure. For example, two dense, high-similarity clusters connected by a single edge likely represent two distinct entities, but transitive closure would incorrectly merge them into one. The risk of falsely combining clusters increases with population size; hence, information on graph structure may be particularly important for linkage accuracy at scale.

To address this limitation, clustering approaches have been developed that leverage network structure to more accurately resolve entities. These methods fall broadly into two categories[5]. The first set of approaches directly models the linkage problem as a bipartite graph linking observed records to latent entities[7–10], explicitly accounting for uncertainty in cluster assignment. While these approaches offer principled uncertainty quantification, they are computationally intensive and often do not scale well to large administrative datasets. The second category treats clustering as a post-processing step: pairwise similarity scores are first generated for each record pair (e.g., using the Fellegi-Sunter framework[3]), producing a weighted similarity graph. Graph-based clustering algorithms are then applied to partition records into clusters representing unique individuals[11–14]. These approaches are typically computationally efficient and easily parallelizable, thus better suited to large-scale administrative data settings. We will restrict our attention to graph-based post-processing approaches.

### Related work in graph-based post-processing

One of the earliest graph-based post processing approaches was proposed by Monge and Elkan[15], who proposed a union-find data structure to compute transitive closure incrementally, and priority queues to reduce expensive pairwise comparisons. A number of clustering approaches for duplicate detection have since been developed. A 2009 review by Hassanzadeh et al.[12] provided a comprehensive comparison of unconstrained clustering algorithms and identified center, merge-center, and Markov clustering as particularly effective in terms of scalability and ability recover true clusters. More recently, Wang et al.[16] proposed the G-Cluster and H-Cluster algorithms, demonstrating their efficiency in large-cluster settings. Draisbach et al.[17] subsequently introduced the extended maximum clique clustering (EMCC) approach and provided a thorough comparative evaluation, showing that EMCC and Markov clustering perform particularly well when clusters are large.

A large body of graph clustering algorithms has also been developed in related fields such as network community detection and information theory[18–25]. Despite being conceptually aligned with entity resolution – both aim to partition nodes into clusters with strong within-cluster and weak between-cluster similarity – these methods have seen limited application in entity resolution. Several such approaches, including Louvain[19] and Leiden[24], have demonstrated strong empirical performance, scale efficiently to large graphs, and are readily implementable using existing software packages. To our knowledge, only one study has applied Louvain to entity resolution[26], and while its results are promising, no systematic evaluation against established methods exists. This presents an important gap given the potential of graph clustering algorithms as scalable alternatives to improve entity resolution in large administrative datasets.

### Adaptation of clustering algorithms to entity resolution settings

Many clustering algorithms were not designed to account for the unique structural properties of record linkage graphs. In entity resolution, edge density often reflects the volume of pairwise comparisons made rather than the strength of evidence for a true match. For example, a common first name such as “John” will generate many candidate pairs with low discriminative value. Conversely, sparse but highly informative edges can provide strong evidence of a match – for example, a rule designed to capture nickname pairs (e.g., Margaret and Peggy) may produce few edges, but each represents strong evidence that candidate pair refer to the same individual. Algorithms not designed for this setting may incorrectly merge clusters connected only by many weak edges, treating their cumulative weight as evidence of a match, or fail to merge clusters linked by a small number of strong edges, underweighting their discriminative value, both of which reduce linkage quality.

In this paper, we conduct a systematic comparison of graph-based clustering approaches for the post-processing step in entity resolution, evaluating performance across datasets of varying size and complexity. We consider established clustering approaches for entity resolution as well as approaches from related domains that have not yet been applied but show potential. We focus on algorithms that are readily implemented via existing software and can scale to datasets with millions of records. We additionally propose a simple adaptation to address the structural properties described above: we (i) remove weakly informative edges prior to clustering to reduce noise, and (ii) after clustering, merge disconnected clusters linked by sufficiently strong edges. We evaluate the proposed adaptation through both illustrative examples and systematic evaluations.

## SETUP

Let $\mathcal{R} = \{r_1, \dots, r_m\}$ denote a set of $m$ records belonging to $n \leq m$ distinct individuals. The goal of entity resolution is to partition $\mathcal{R}$ into $K = n$ disjoint clusters $C = \{c_1, \dots, c_K\}$, where $K$ is unknown, such that $\bigcup_{l=1}^{K} c_l = \mathcal{R}$ and $c_l \cap c_{l'} = \emptyset$ for all $l \neq l'$.[12,17] Each cluster $c_l$ contains all records corresponding to a single unique individual.

Since exhaustive pairwise comparison over all $\binom{m}{2}$ candidate pairs are computationally prohibitive for large $m$, blocking is typically employed to restrict comparisons to pairs sharing common attributes (e.g., identical first name or date of birth), yielding a candidate set $\mathcal{P} \subseteq \binom{\mathcal{R}}{2}$. Each candidate pair $(r_i, r_j) \in \mathcal{P}$ is then assigned a similarity score $s_{ij} \in [0,1]$, with higher score representing higher similarity. The score $s_{ij}$ may be derived from probabilistic frameworks such as Fellegi-Sunter (where $s_{ij}$ is based on a mixture model of scores for true matches and true non-matches) or from other similarity measures. In this paper, $s_{ij}$ is obtained via the Fellegi-Sunter framework[3,27,28].

The scored candidate pairs can be represented as a weighted undirected graph $G$, where each node corresponds to a record $r_i$, each edge corresponds to a candidate pair $(r_i, r_j) \in \mathcal{P}$, and the weight $w(r_i, r_j) = s_{ij}$ reflects the similarity of the two records.

### Criteria for inclusion of graph clustering algorithms

Clustering algorithms were considered for inclusion as a post-processing approach if they satisfied the following criteria:

1. **Takes graph as input**. The algorithm operates directly on the graph $G$, which may be weighted or unweighted. In the unweighted case, a threshold $\theta$ is applied such that an edge $(r_i, r_j)$ is retained if $s_{ij} \geq \theta$, and the weights $s_{ij}$ are not used to inform clustering. Algorithms that do not operate on graphs (e.g., require records to be represented in a Euclidean space) were excluded.
2. **Partitioning into disjoint clusters.** The algorithm returns a partition of $\mathcal{R}$ into disjoint clusters. Algorithms returning overlapping or soft cluster assignments are excluded.
3. **Code availability.** The algorithm must satisfy at least one of the following: (a) there is a maintained Python implementation that is publicly available, (b) the algorithm is described by pseudocode that is sufficiently detailed and straightforward to implement, or (c) code was obtained directly from authors upon request.
4. **Scalability.** We included clustering algorithms described as scalable in prior systematic evaluations[12,17] or by their original authors (sources in Table 1). Scalability has no standard definition in this literature and has been evaluated using different criteria, graph sizes, and computational environments. We therefore deferred to each source's assessment of scalability.
5. **No inestimable input parameters**. The algorithm does not require parameters that cannot be determined prior to clustering. For example, algorithms requiring the number of clusters $k$ as input were excluded, since this quantity is unknown in practice. Tuning parameters are allowed, as it can be tuned on labeled data without knowing the true cluster structure.

## SELECTED CLUSTERING ALGORITHMS

Eleven clustering algorithms met the above inclusion criteria and were tested as post-processing methods for entity resolution (Table 1). Five clustering algorithms had been previously systematically evaluated in entity resolution: transitive closure, center[29], merge-center[30], Markov clustering[31], and weighted diameter[32]. The remaining six algorithms satisfied the inclusion criteria but had not been previously systematically evaluated in an entity resolution context: Leiden (with modularity or constant Potts model (CPM) objective function)[24], Louvain[19], Infomap[22,23], label propagation[21], fast greedy[25], and walktrap[20].

Table 1 summarizes the key differences among these algorithms. Stochastic algorithms involve randomized steps in the clustering process and may yield different partitions across runs, while deterministic algorithms return the same partition each run. A few algorithms require tuning parameters (e.g., resolution parameter or similarity threshold), which require ground truth labels when tuned for accuracy. Table 1 also reports on each algorithm's time complexity and code availability.

A detailed description of all algorithms included is provided in the appendix (under "Description of clustering algorithms"). A comprehensive evaluation for all algorithms considered against the inclusion and exclusion criteria is provided in Appendix **(Table S6).**

**Table 1.** List of clustering algorithms meeting all inclusion criteria ($N$: number of vertices, $E$: number of edges).

| Algorithm | Stochastic or deterministic | Tuning parameters | Time complexity | Code availability |
|---|---|---|---|---|
| Transitive closure | Deterministic | Threshold $\theta$ | $\mathcal{O}(1)$ | Pseudocode |
| Leiden (modularity)[24] | Stochastic | Resolution $\gamma$ | $\mathcal{O}(N \log N)$ | iGraph package |
| Leiden (CPM)[24] | Stochastic | Resolution $\gamma$ | $\mathcal{O}(N \log N)$ | iGraph package |
| Louvain[19] | Stochastic | Resolution $\gamma$ | $\mathcal{O}(N \log N)$ | iGraph package |
| Infomap[22,23] | Stochastic | None | $\mathcal{O}(E)$ | iGraph package |
| Label propagation[21] | Stochastic | None | $\mathcal{O}(E)$ | iGraph package |
| Fast greedy[25] | Deterministic | None | $\mathcal{O}(E\ d \log N)$<br>$d$: depth of dendogram<br><br>$\mathcal{O}(N \log^2 N)$ on sparse graphs | iGraph package |
| Walktrap[20] | Stochastic | Steps $t$ | $\mathcal{O}(N^2 \log N)$ to $\mathcal{O}(EN^2)$ | iGraph package |
| Center[29] | Deterministic | None | $\mathcal{O}(E \log(E))$ | Pseudocode |
| Merge-center[30] | Deterministic | None | $\mathcal{O}(E \log(E) + E)$ | Pseudocode |
| Markov[31] | Deterministic | Inflation $r$ | $\mathcal{O}(Nk^2)$<br>$k$: pruning constant | markov_ clustering package |
| Weighted diameter[32] | Stochastic | Distance $d$ | Unknown | Code from authors |

**See Appendix Table S6 for comprehensive evaluation for all algorithms considered against the inclusion and exclusion criteria.*

## ADAPTATION OF CLUSTERING ALGORITHMS TO ENTITY RESOLUTION

To account for the structural properties of record linkage graphs described in the introduction, we propose the following adaptation, applied uniformly across all clustering algorithms evaluated **(Appendix Figure S1)**:

1) Prior to clustering, edges with $s_{ij} < \theta_{low}$ are removed from the graph $G$ to reduce noise from weakly informative comparisons, yielding a filtered graph $G_{\theta_{low}}$.
2) The clustering algorithm is applied to $G_{\theta_{low}}$ to obtain an initial partition.
3) As a post-processing step, any two clusters that are disconnected in $G_{\theta_{low}}$ but linked by at least one edge with $s_{ij} \geq \theta_{high}$ in the original graph $G$ are merged, ensuring that strong candidates of true matches record pairs are not discarded.

The selection of $\theta_{low}$ and $\theta_{high}$ are described under "optimization of tuning parameters".

## OPTIMIZATION OF TUNING PARAMETERS

Tuning parameters for each clustering algorithm (threshold $\theta$, resolution $\gamma$, steps $t$, inflation $r$, or distance $d$) were selected to maximize linkage accuracy (i.e., F-score, defined in "accuracy measures"). Optimization was done using a coarse-to-fine search: parameters were first evaluated over a coarse range, the region yielding best F-score was identified, and the search was then refined within the neighborhood of that region until no further improvement was obtained.

For clustering with adaptation to entity resolution settings, adaptation parameters ($\theta_{low}$ and $\theta_{high}$) were optimized jointly with the other tuning parameters using the same procedure. The optimal tuning parameter values were reported in **Appendix Tables S1-S4**.

## EXPERIMENTAL EVALUATION

### Implementation

We evaluated the performance of clustering algorithms on several synthetic entity resolution datasets and one real-world dataset from the linkage of the South Africa National Health Laboratory Service (NHLS) National HIV Cohort[33]. We describe the data source and the fields used in pairwise comparison below; details of the blocking and scoring strategies used to generate edge weights in the similarity graph in the simulated data can be found on GitHub.

Probabilistic record linkage for the synthetic datasets was performed using Splink[6], which was used to implement blocking, generate candidate pairs, and score pairwise comparisons based on the Fellegi-Sunter model. Probabilistic linkage of the NHLS HIV Cohort was done in-house at the National Institute of Communicable Diseases (NICD), South Africa[34]. In both cases, record linkage produced a scored edge list, where each edge represents a candidate pair and weighted by its match probability. Prior to clustering, edges with a match probability below 0.05 were discarded as an initial cleaning step. (These edges represent near certain non-matches whose inclusion would introduce noise and increase computational burden). Clustering algorithms were then applied directly to the resulting edge list.

All analyses were conducted in Python (version 3.10.12). Python was chosen due to its availability of well-performing and actively maintained record linkage libraries (e.g., Splink), enabling the integration of clustering algorithms to existing record linkage pipeline as a post-processing step. Most clustering algorithms were implemented using iGraph (version 1.0.0);[35] Markov clustering was implemented using the Markov_clustering library (version 0.0.6)[36], center and merge-center were implemented from pseudocode, and the weighed diameter algorithm was implemented from code obtained from the original authors.[32]

Experiments were run on the shared computing cluster (SCC) at Boston University with 4 cores and 32 GB of RAM.

**Datasets**

**FEBRL3.** FEBRL3 is a realistic synthetic dataset generated using the Freely Extensible Biomedical Record Linkage (FEBRL) data generator[37] and have been widely used as a benchmark dataset in record linkage. It is available as a built-in dataset in Splink[6]. The frequency and error distribution of attribute values are modelled to reflect real-world data characteristics. The dataset contains 5,000 records, with 2,000 originals and 3,000 duplicates. Duplicates follow a Zipf distribution with a maximum of five duplicates per original. Available fields include first name, surname, street number, address, suburb, postcode, state, date of birth, and social security ID. Each record includes a true cluster identifier for performance evaluation.

**Historical 50K.** The Historical 50K dataset is a synthetic benchmark dataset available as a built-in dataset in Splink[6]. Records are based on historical persons sourced from Wikidata, with duplicates generated by introducing a variety of errors into the originals. The dataset contains 50,000 records, with 5,156 originals and 44,844 duplicates. Fields include first name, surname, date of birth, birthplace, postcode, gender, and occupation, and each record includes a true cluster identifier for performance evaluation.

**Synthetic 7M.** The Synthetic 7M dataset was generated using open-source synthetic data generation code published by the UK Ministry of Justice.[6] As with the Historical 50K dataset, records are based on historical persons sourced from Wikidata. Each original record was corrupted to produce between 0 and 35 duplicates following a Zipf distribution, with a median of 8 duplicates per original. The dataset contains 7 million records, including 796,504 originals and 6,203,496 duplicates. Fields include first name, middle name, surname, birth location (latitude, longitude, and country), occupation, country of citizenship, date of birth, and date of death. Each record includes a true cluster identifier for evaluation. The dataset is not publicly available but a similar dataset can be reproduced using open-source code released in the Splink GitHub repository.[6]

**NHLS HIV cohort.** The South Africa NHLS National HIV Cohort contains over 2 billion HIV and HIV-related laboratory test records from 1 January 2004 to 31 December 2023[32,33]. The dataset includes all individuals receiving care in the South African public health sector, which serves over 80% of the country's population. As individuals may seek care repeatedly over time, the dataset may contain multiple laboratory records per individual. However, approximately 85% of the records lack a unique identifier, and missing values and typographical errors are common across identifying fields (first name, surname, date of birth, gender, care facility, and province), making follow-up difficult. Entity resolution is therefore essential to cluster records belonging to the same individual and enable longitudinal tracking of patient outcomes.

For confidentiality reasons, we work with a fully de-identified edge list with pairwise match scores derived from a Fellegi-Sunter framework, previously linked at NICD[34]. For the purposes of this study, clustering algorithms are evaluated on a subset of 6.9 million unique records drawn from the full edge list. Ethics approval for analysis of de-identified NHLS data was obtained from University of Witwatersrand Human Research Ethics Committee (M200447), Boston University Medical Campus Institutional Review Board (H-31968), and NHLS Academic Affairs and Research Management System Review Committee (PR2010539).

## Accuracy measures

We evaluate the performance of the clustering algorithms using three standard measures: positive predictive value (PPV, i.e., precision), sensitivity (SEN, i.e., recall), and F-score. PPV measures the proportion of true matches out of all predicted matches, SEN measures the proportion of true matches successfully identified by clustering, and F-score measures the harmonic mean between PPV and SEN. For each clustering algorithm, metrics are reported at tuning parameters that maximizes the F-score (see "Optimization of tuning parameters").

A naive "all pairwise" evaluation, i.e., comparing all possible record pairs, is biased toward large clusters, since a cluster of $n$ nodes contribute $\binom{n}{2}$ pairs for evaluation (quadratic in $n$). For example, 100 nodes linked together into a single cluster contribute 4,950 pairs while 100 nodes linked into 50 clusters of size 2 contribute only 50 pairs. Thus, the metric is sensitive to errors in large clusters while being relatively insensitive to errors in small clusters.

We instead adopt a "reference node" approach. For each cluster, we assign one node as the reference node, and all other nodes in the cluster are compared against the reference node. This ensures each node contributes equally to the evaluation, regardless of cluster size. A misassigned node counts as one error whether it occurs in a cluster of size 2 or 100.

The approach proceeds as follows:

Let $G = \{G_1, \dots, G_K\}$ denote the set of ground truth clusters and $C = \{C_1, \dots, C_{K'}\}$ the set of predicted clusters identified by the clustering algorithm.

To calculate SEN, a reference node $r_k$ is selected at random for each ground truth cluster $G_k$. For each non-reference node $i \in G_k \backslash \{r_k\}$, the pair $\{r_k, i\}$ is a true positive ($TP_{r_k,i}$) if $r_k$ and $i$ are assigned to the same predicted cluster in $C$, and a false negative ($FN_{r_k,i}$) otherwise. Hence,

$$SEN = \frac{\sum_k \sum_{i \in G_k \backslash \{r_k\}} TP_{r_k,i}}{\sum_k \sum_{i \in G_k \backslash \{r_k\}} TP_{r_k,i} + \sum_k \sum_{i \in G_k \backslash \{r_k\}} FN_{r_k,i}}. \tag{1}$$

To calculate PPV, a reference node $r_{k'}$ is selected at random for each predicted cluster $C_{k'}$. For each non-reference node $i \in C_{k'} \backslash \{r_{k'}\}$, the pair $\{r_{k'}, i\}$ is a true positive ($TP_{r_{k'},i}$) if $r_{k'}$ and $i$ belong to the same ground truth cluster in $G$, and a false positive ($FP_{r_k,i}$) otherwise. Hence,

$$PPV = \frac{\sum_{k'} \sum_{i \in C_{k'} \backslash \{r_{k'}\}} TP_{r_{k'},i}}{\sum_{k'} \sum_{i \in C_{k'} \backslash \{r_{k'}\}} TP_{r_{k'},i} + \sum_{k'} \sum_{i \in C_{k'} \backslash \{r_{k'}\}} FP_{r_{k'},i}} \tag{2}$$

and

$$Fscore = \frac{2 \cdot PPV \cdot SEN}{PPV + SEN}. \tag{3}$$

We note that this approach differs from the "all pairwise" evaluation used in prior studies; we prefer it because each node contributes equal weight to the metric irrespective of cluster size. To account for variability due to the choice of reference node, all results are averaged across 5 independent runs. We also note that the F-score reported here differs from the traditional pairwise F-score: SEN and PPV are computed over different sets of clusters, and their true positive counts therefore generally differ. Thus, F-score cannot be expressed as 2TP/(2TP+FP+FN).

For the synthetic datasets, ground truth information is available for all records, allowing direct evaluation against the complete ground truth without correction. For the NHLS HIV cohort, where ground truth is available for only 15% of records, a bias-correction approach is applied to account for the non-representativeness of this subset[38].

**Robustness checks**

Following prior work[12,16,17], we report within-sample performance for our primary estimates. To investigate out-of-sample performance and the stability of these estimates, we additionally evaluated performance on the NHLS HIV cohort using a separate test subset of 6.9 million records drawn from the full edge list that was not used for tuning. Results are reported in **Appendix Table S5**.

Several of the graph-clustering algorithms are stochastic and may produce different partitions across runs due to randomized steps in the clustering process. To assess the stability of clusters, we repeated the clustering step for the top-performing algorithms across five runs and classified each cluster as stable if it was unchanged across all five runs and unstable otherwise. Results are reported in **Appendix Table S7**.

**An upper bound on clustering performance**

Following Draisbach et al.[17], we establish an upper bound on clustering performance, in which records are assigned to clusters using their ground truth cluster identifiers, restricted to record pairs retained after blocking. Since only true matches are assigned to the same cluster, there are no false positives and PPV is always 100% by construction. However, SEN may be less than 100%, reflecting true edges excluded at the blocking step, i.e., pairs that were never compared and therefore cannot be recovered by any clustering algorithm (**Figure 1**). This upper bound provides a useful benchmark for the maximum achievable performance for any clustering algorithm given the blocking strategy used. Estimating this upper bound requires the ground truth cluster identifiers that entity resolution is intended to recover and thus cannot be applied in practice; it serves only as a theoretical point for comparison.

**Figure 1.** Blocking, not clustering, sets the upper bound on sensitivity. A true match dropped at the blocking stage cannot be recovered by any clustering algorithm. This sets the upper bound on clustering performance, described above.

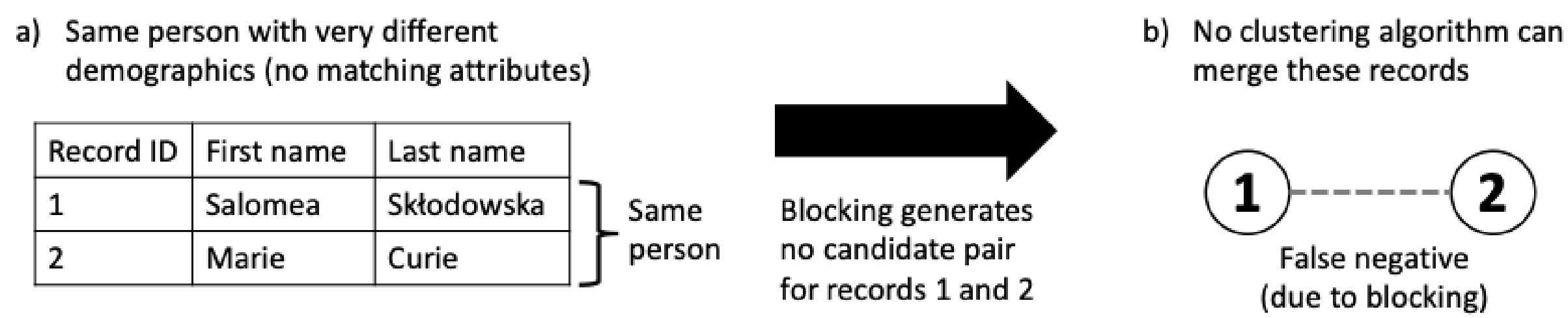

### Percent reduction in linkage error

Gaps in linkage performance are estimated as $1 - PPV$ (overmatching, false discovery rate), $1 - SEN$ (undermatching, false negative rate), and $1 - F\text{-}Score$, which we define as a catch-all "linkage error rate". Benchmarking against the upper bound, we estimate "percent reduction in linkage error" relative to transitive closure for each clustering algorithm $a$ as follows:

$$\%\ reduction\ in\ overmatching_a = 100\% - 100\% \cdot \frac{PPV_{UpperBound} - PPV_a}{PPV_{UpperBound} - PPV_{TransitiveClosure}}, (4)$$

$$\%\ reduction\ in\ undermatching_a = 100\% - 100\% \cdot \frac{SEN_{UpperBound} - SEN_a}{SEN_{UpperBound} - SEN_{TransitiveClosure}}, (5)$$

$$\%\ reduction\ in\ linkage\ error_a = 100\% - 100\% \cdot \frac{F_{UpperBound} - F_a}{F_{UpperBound} - F_{TransitiveClosure}}. \quad (6)$$

## RESULTS

### Evaluation of clustering performance

**Figure 2** shows validation performance for each algorithm at its best associated tuning parameters, without adaptation to entity resolution settings, across all datasets evaluated. **Figure 3** shows the corresponding results with adaptation. In each figure, the top row (upper bound) indicates the best PPV and SEN attainable: PPV is always 100%, but SEN may fall below 100% because some edges are lost during blocking. Note that for the NHLS HIV cohort data, the gold standard is imperfect – multiple national IDs may be associated with records sharing identical identifying information, hence PPV is not 100% in this case. We compare all graph clustering algorithms against transitive closure, which is the current standard practice.

Throughout, we rank all algorithms by the F-score, and unless otherwise stated all reported percentages in the text refer to performance after adaptation to entity resolution settings.

FEBRL3 is a small dataset, and transitive closure alone already achieves very good performance **(Figure 3, Appendix Table S1).** Applying the clustering algorithms yields modest improvements. Leiden with the CPM objective function performs best, reducing the false negative rate by 54.6% relative to transitive closure (a 0.8% increase in SEN), followed closely with Markov clustering and center. Out of the top performing algorithms, Markov is the slowest with a runtime of 3 seconds compared to 35 milliseconds for Leiden CPM. The f-score for the top performers is largely unchanged with and without adaptation.

On Historical 50k, a medium-sized dataset, the clustering algorithms produce larger improvements over transitive closure **(Figure 3, Appendix Table S2)**. Leiden modularity performs best, leading to an 88.2% reduction in false negative rates (a 9.7% increase in SEN) and a 15.9% reduction in false discovery rates (0.9% increase in PPV) relative to transitive closure, equivalent to a 65.9% reduction in linkage error rates, followed closely with Louvain and label propagation. Label propagation has the additional advantage of being fully unsupervised, requiring no tuning parameters to be set. All three algorithms are fast, with runtimes of 0.5-1 seconds in our computing environment (see experimental evaluation). Adaptation leaves the f-score for the top performers largely unchanged.

For the larger datasets (Synthetic 7m and NHLS HIV cohort), we evaluated all algorithms that were computationally feasible. Walktrap, with quadratic to near-cubic time complexity, did not complete within 72 hours on our computational environment and was omitted.

For the Synthetic 7m dataset, Leiden modularity performs best, followed closely by Louvain and label propagation **(Figure 3, Appendix Table S3)**. Leiden modularity and Louvain reduce the false negative rates by 62.5% (a 2.5% increase in SEN) and the false discovery rate by 30.6-31.3% relative to transitive closure (a 0.9% increase in PPV), equivalent to a 49.6-49.9% reduction in linkage error rates. Label propagation reduces the false negative rate by 49.5% and the false discovery rate by 46.3%. Runtimes range from 2.5 minutes for the Leiden and Louvain algorithms, and 6 minutes for label propagation. Adaptation led to modest improvements in f-score values (0.3-0.4%).

For the NHLS HIV cohort data, Leiden CPM performs best, reducing the false negative rates by 58.2% (2.5% increase in SEN) and the false discovery rate by 17.5% (a 0.8% increase in PPV) relative to transitive closure (a 36.8% reduction in linkage error rate), closely followed by Louvain and Leiden modularity **(Figure 3, Appendix Table S4)**. Adaptation improves the f-score by 0.5% for Leiden CPM, and by 2.7-2.8% for Louvain and Leiden modularity. Notably, after adaptation, Louvain, Leiden CPM, Leiden modularity, and label propagation outperform the weighted diameter algorithm with substantial improvements in runtime: 2-3 minutes for Louvain and Leiden and 9.5 minutes for label propagation compared to 1 hour for weighted diameter.

Across all datasets, fast greedy, merge-center, and walktrap did not provide any notable performance improvements over transitive closure. Center is fast and more accurate than transitive closure, but falls behind the top performing algorithms (Leiden, Louvain, label propagation).

On the out-of-sample NHLS HIV cohort test subset, the F-score differed from the within-sample estimate by 0.1% or less, suggesting minimal overfitting **(Appendix Table S4**). Assessing the stochasticity of the clusters, Leiden was the most stable, with 99.3% of clusters unchanged across five runs under CPM and 99.5% under modularity. Louvain was slightly less stable, with 98.7% of clusters remaining stable. Label propagation is the least stable of the three, with 95.3% of clusters remaining stable across five runs **(Appendix Table S7).**

**Figure 2.** Validation performance, without adaptation to entity resolution settings, sorted by descending order in f-score. Point estimates show means across five runs; error bars show 2.5th-97.5th percentiles. Vertical dotted line indicates the upper bound; algorithms below the horizontal dotted line underperform transitive closure.

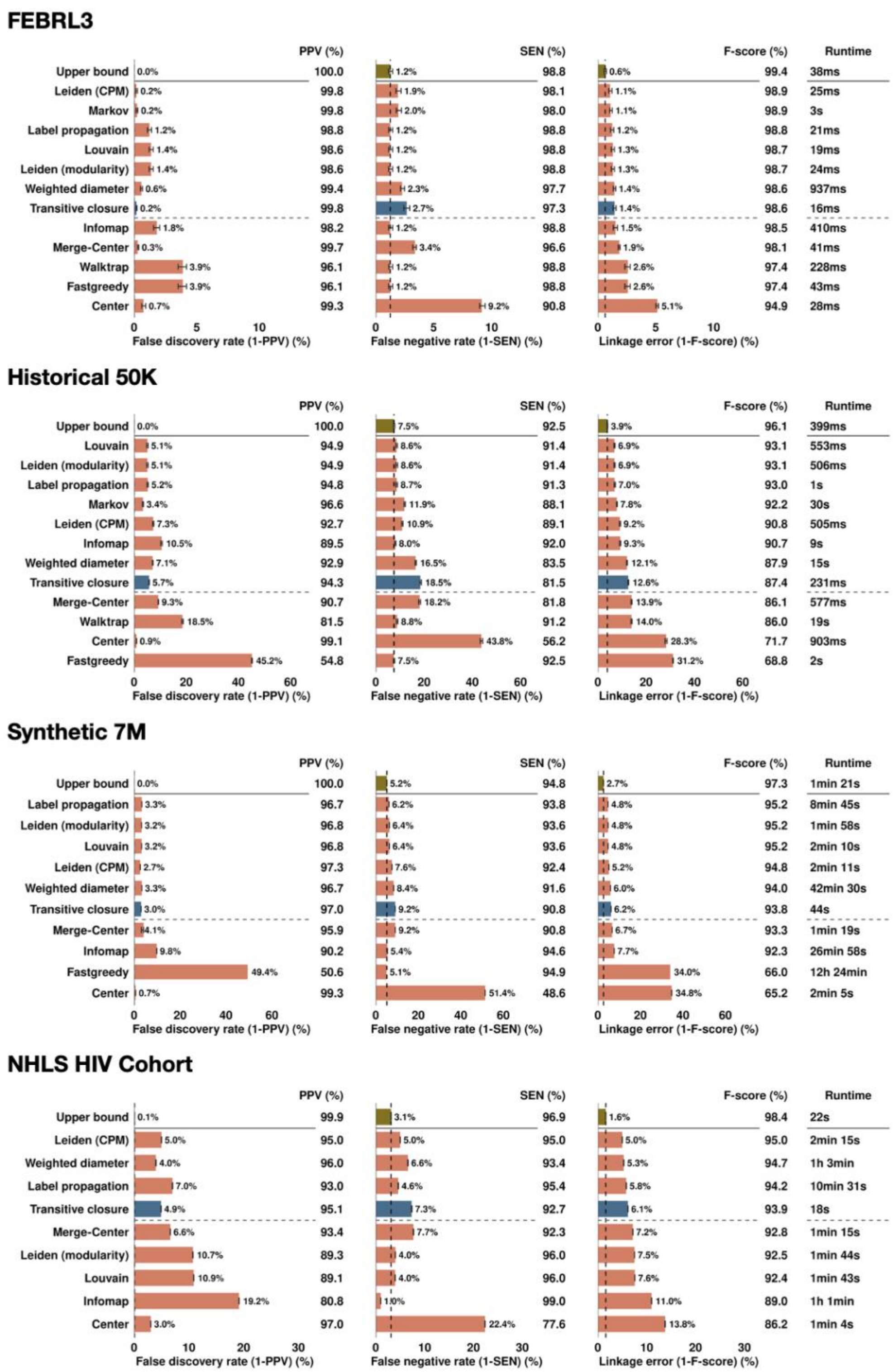

**Figure 3.** Validation performance, with adaptation to entity resolution settings, sorted by descending order in f-score. Point estimates show means across five runs; error bars show 2.5th-97.5th percentiles. Vertical dotted line indicates the upper bound; algorithms below the horizontal dotted line underperform transitive closure.

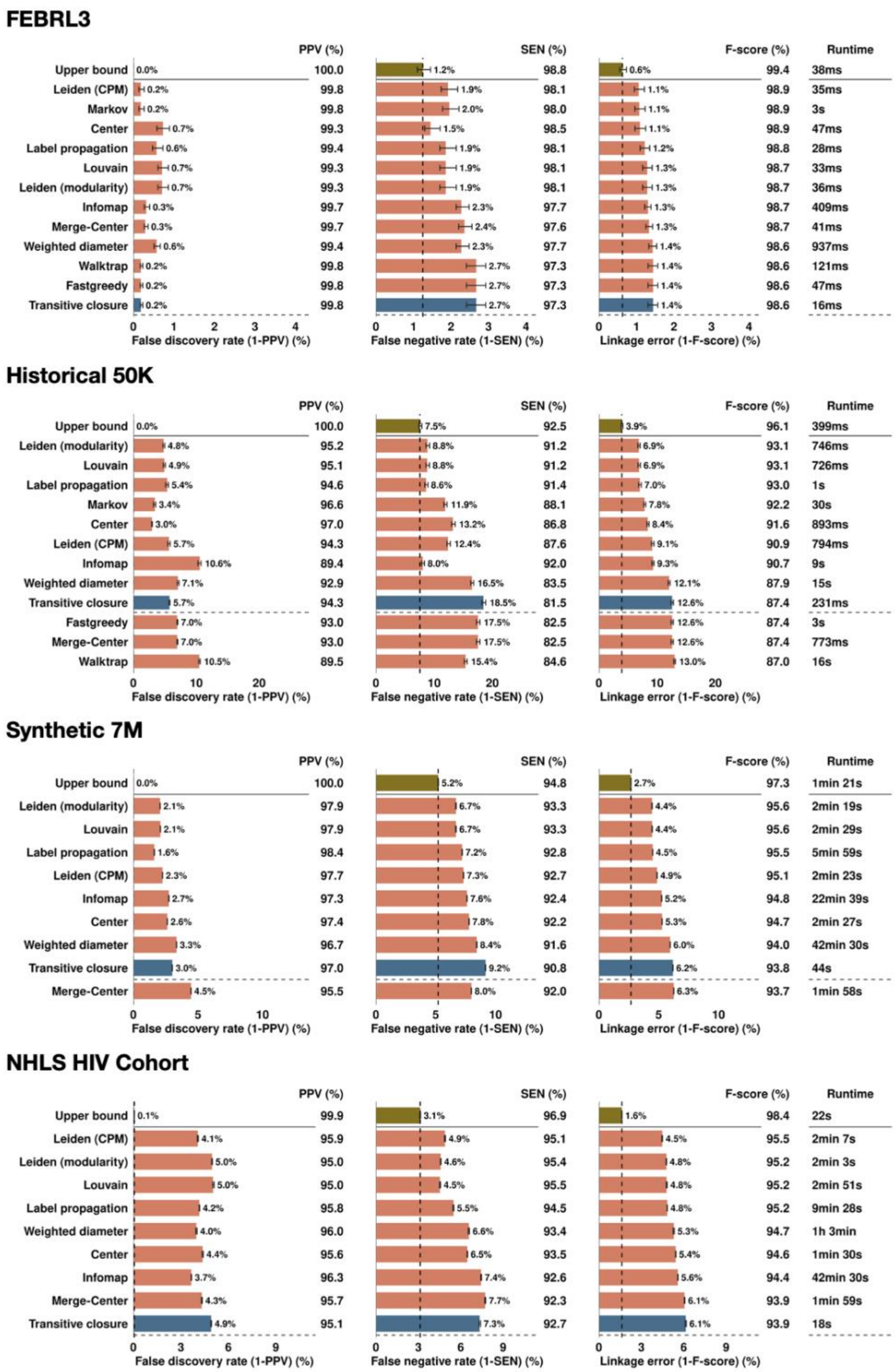

## MECHANISMS OF IMPROVEMENT OVER TRANSITIVE CLOSURE

### How top-performing algorithms assign cluster membership

All three best-performing graph clustering algorithms – Leiden, Louvain, and label propagation – assign cluster membership by weighing each edge against the surrounding cluster structure (e.g., the density and connectivity of a node's neighborhood), rather than treating each edge in isolation. This improved PPV and SEN over transitive closure, which applies a fixed cutoff to each edge independently **(Appendix Figure S2 a)**. The three approaches differ in how they use information on the cluster structure:

Louvain and Leiden share a similar procedure: each node starts in its own cluster and is moved to the neighboring cluster that provides the largest positive gain in the objective function (local moving of nodes phase) **(Appendix Figure S2 b)**[19,24]. The objective function rewards high within-cluster density and low between-cluster density; lower-scoring edges are tolerated within dense or small clusters, while a sparse bridge between two large dense clusters may be dropped. All nodes in the same cluster are then aggregated into a single node (aggregation phase), and the process repeats until no further improvement is possible. The difference is that Leiden adds a refinement step guaranteeing well-connected clusters, correcting a known issue in Louvain, where clusters may be poorly connected or even disconnected[24]. In practice, the two performed very similarly in our data.

Louvain aims to maximize the modularity[39] objective function, while Leiden can maximize either modularity or the constant Potts model[40] (CPM). Modularity maximizes the difference between the between the total weight of within-cluster edges and what was expected under a degree-matched random graph. As modularity depends on some global parameters, it is known to be subject to resolution limit, where it is unable to detect small clusters within a large graph[41]. CPM works by maximizing the total weight of within-cluster edges, where a cluster is retained only when its internal density exceeds the resolution $\gamma$ and is split otherwise; it is resolution limit free by construction as it uses local weights[40]. The multi-level procedure of Louvain and Leiden also appears to mitigate the resolution limit issue of modularity, as noted by Blondel et al.[19] The first phase move nodes individually from one cluster to another, hence two distinct clusters are unlikely to merge. These clusters may still be merged in later passes after aggregation.

Label propagation begins with each node holding a unique label; each node then iteratively adopts the label held by the majority of its immediate neighbors **(Appendix Figure S2 c)**[21]. The process repeats until every node holds a label shared by the maximum number of its neighbors, and nodes that share the same label form a cluster. Clustering is thus performed locally, by consensus among immediate neighbors, requiring no objective functions or tuning parameters.

The original implementation by Raghavan et al. operates on an unweighted graph; the iGraph implementation used in this paper extends this by allowing each neighbor's label to be weighted by its edge weight.[42]

Note that Louvain, Leiden, and label propagation are stochastic and may yield different partition across runs. In Louvain and Leiden, the order in which nodes are visited during the "local moving" phase is randomized. Different orderings may lead to different local optima. In label propagation, ties (i.e., two or more labels are equally common among a node's neighbors) are broken at random. We assessed cluster stability across repeated runs and found Leiden and Louvain to be highly stable (>98.7% of clusters unchanged across runs) while label propagation showed higher instability (95.3% stable). We recommend fixing a random seed for reproducibility across runs.

**Consequences for linkage accuracy**

Across all datasets, graph clustering led to large improvements in SEN and minor improvements in PPV over transitive closure. The asymmetry reflects how each method handles weak vs. strong edges. The two methods diverge mainly on how low- and mid-scoring (uncertain) edges are handled, which drives sensitivity. Transitive closure drops every edge below its threshold, with no way to recover them. Graph clustering instead evaluates each edge against its neighborhood: a mid-scoring edge inside a dense group, or within a small cluster, may be retained rather than dropped, recovering true matches missed by transitive closure **(Figure 4, left panel)**. The pattern reflects how entity resolution behaves in practice – small clusters often represent individuals with less common attributes, where a mid-scoring link still provides meaningful evidence of a match (since having any link is less common for rare individuals). Likewise, dense clusters represent individuals with many overlapping identifiers.

Both approaches handle high-scoring edges similarly: transitive closure always retain edges above its threshold, keeping those clusters intact. Likewise, high-scoring edges tend to lie within dense clusters that is kept intact by graph clustering algorithms. Because transitive closure already achieves high PPV (i.e., most edges exceeding the threshold are true positives), few false positives remain to correct, and PPV improves only marginally. Graph clustering may improve PPV in the case where two large, dense clusters are joined by a single edge **(Figure 4, right panel)**. Such clustering structure likely represents two distinct individuals with a single spurious link. As the spurious link is weak relative to each sub-cluster's internal density, graph clustering splits the cluster apart.

Applying graph clustering to entity resolution graphs raises two considerations. First, graph clustering can place too much weight on a group of low-scoring edges when they form a dense structure, treating density as evidence of a match. In practice, low-scoring edges mostly represent noise. Second, a single high-scoring link represents a near certain match but might be overlooked by graph clustering due to lack of density. Our proposed adaptation addresses both: dropping low-scoring edges prior to clustering and merging back very high-scoring edges after clustering further improved performance without substantially affecting runtime.

**Figure 4.** An illustrative example where graph-based clustering provides improvement over transitive closure, in the NHLS HIV cohort. The figure shows four connected components of the linkage graph (A-D), partitioned with transitive closure in the top row and Leiden (CPM) with adaptation in the bottom row using the optimal tuning parameters. (Left panel (components A and B) shows improvement in SEN; right panel (components C and D) shows improvement in PPV). Node color denotes entity based on national ID (ground truth) data: blue (entity 1), orange (entity 2), and grey (unknown entity). Edge style denotes linkage status by clustering algorithm (solid line: linked, dashed line: unlinked).

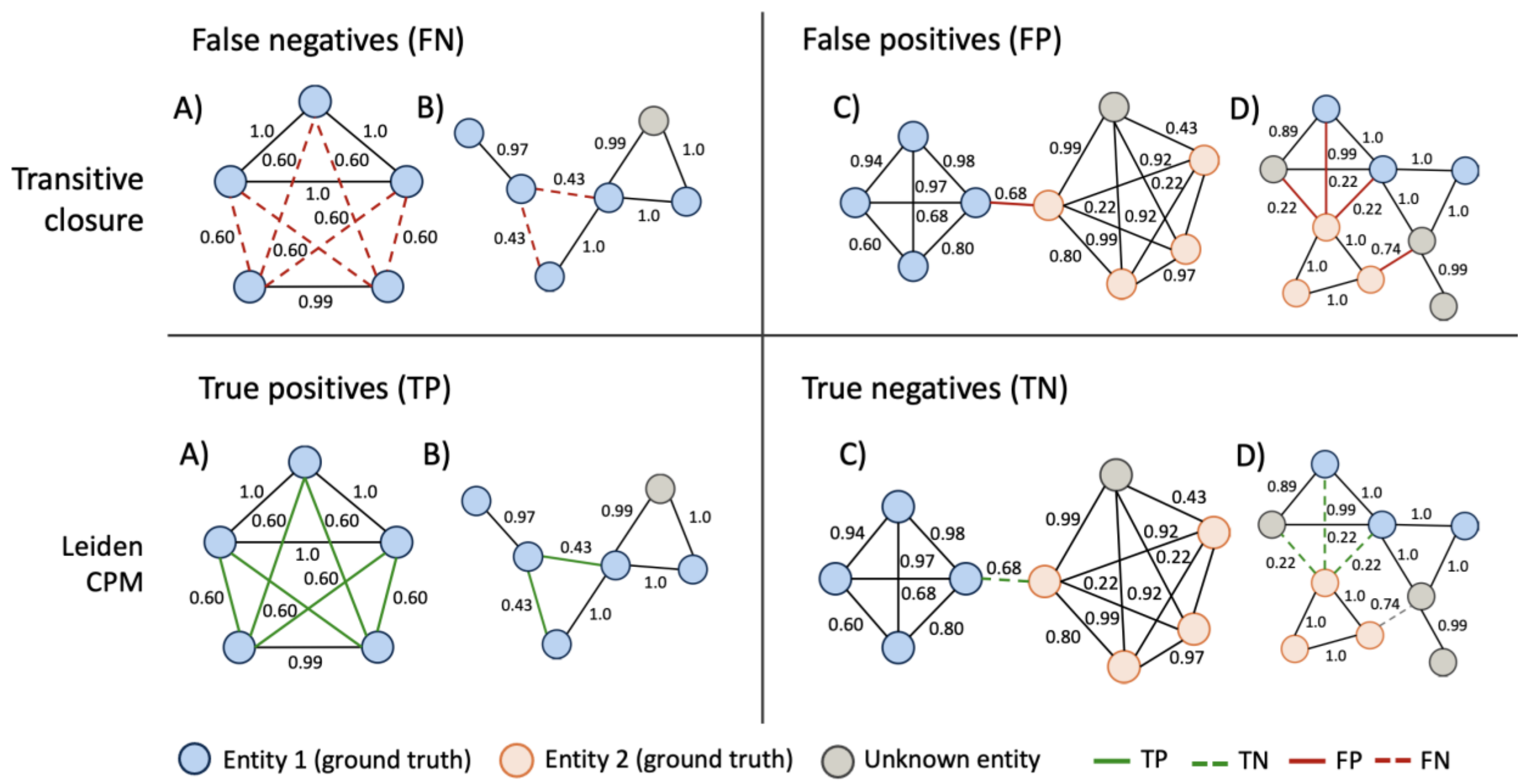

## DISCUSSION

To our knowledge, this is the first paper to systematically evaluate graph-based clustering algorithms from adjacent fields (community detection and information theory) against established clustering algorithms from the entity resolution literature. We found that several of these cross-field algorithms, namely Leiden, Louvain, and label propagation, achieved comparable or better linkage accuracy than existing entity resolution approaches. These cross-field algorithms also offer several practical advantages: they run in near-linear time and are readily implementable in well-maintained software packages.

Prior systematic comparisons have provided in-depth evaluations of existing entity resolution clustering algorithms and repeatedly identified Markov clustering as one of the best performing approaches[12,17]. In line with these findings, Markov achieved the best accuracy among the established algorithms we evaluated. However, it was considerably slower than Louvain and Leiden: on a dataset of 50,000 records, Markov ran in 30 seconds, while Louvain and Leiden ran in 0.7 seconds – a roughly 40-fold speedup – with comparable or better accuracy. Although Louvain and Leiden are stochastic, they produced relatively stable clusters across runs. The center and merge-center algorithms, previously reported as scalable and accurate[12], were fast, with runtimes similar to Leiden and Louvain. However, they underperformed Markov, Leiden, Louvain, and label propagation in our study, with merge-center offering no improvement over transitive closure across all datasets.

The performance improvements from graph-clustering algorithms arise from how cluster membership were assigned. Rather than evaluating each edge independently against a fixed threshold, graph clustering evaluates each edge against the density and connectivity of its surrounding subgraph. Using information on the cluster structure, graph clustering algorithms can recover true matches that fall below a fixed threshold but fall within well-connected or small, tightly linked clusters, accounting for the large improvements in SEN we observed. On the other hand, transitive closure always discards such edges with no possibility of recovery. Information on cluster structure also enables graph clustering algorithms to separate records that score above the fixed threshold but that the surrounding structure indicates multiple entities, resulting in PPV gains.

Louvain and Leiden require tuning a resolution parameter, which needs ground truth information for at least some subset of nodes. The resolution parameter is also less intuitive than transitive closure's threshold: the threshold corresponds directly to a match probability, while a resolution parameter has no such straightforward interpretation. One approach to tuning is a coarse-to-fine search, as we approached in this paper. Tuning is relatively inexpensive as Louvain and Leiden runs very fast: about 3 minutes per iteration on a dataset of 6.9 million records. In very large datasets, tuning can be done on a subset. When tuning on a subset, we recommend sampling whole initial clusters (the connected components of the pre-clustering edge list) rather than sampling nodes or edges – sampling nodes and taking their clusters oversamples large clusters, biasing tuned parameters toward large-cluster performance. Sampling edges does not preserve cluster structure. A few initial clusters can be very large; we recommend dropping edges below a low cutoff (e.g., 0.05) before sampling to keep cluster sizes reasonable.

When ground truth is unavailable, or when fully unsupervised clustering is preferred, label propagation is particularly attractive: it requires no tuning, runs comparably fast to Leiden and Louvain (about 10 minutes for 6.9 million records), and performed nearly as well. However, label propagation is less stable across runs than Leiden and Louvain.

All graph clustering approaches, including existing entity resolution clustering approaches, require only an edge list with similarity weights and a vertex list of record IDs, plus ground truth labels for a subset of records if the algorithm needs tuning. These are standard outputs of many record linkage processes, including Splink[6], so the methods can be incorporated into existing record linkage pipelines with little additional work. Because clustering runs on each connected component independently, it also scales and parallelizes well to large datasets.

Overall, our results suggest that cross-field algorithms – in particular Leiden, Louvain, and label propagation – are among the best options for clustering as a post-processing step in entity resolution. They were among the most accurate of all methods evaluated, ran considerably faster than Markov clustering, and integrate easily into existing pipelines. This study has several limitations. First, we report within-sample estimates in our primary results, following the approach used in prior systematic comparisons[12,16,17]. This may result in optimistic performance estimates; however, our robustness check indicated little overfitting **(Appendix Table S7)**. Nevertheless, risk of overfitting remains, particularly in settings with limited labeled data. Second, we selected tuning parameters via a coarse-to-fine search, which is efficient but explores only a limited region of the parameter space and may fail to find the global optimum. Third, reported runtimes reflect our computational environment and are intended only to provide an estimate of the time required to run each algorithm. Values may vary with hardware, software, and implementations, so they should not be taken as definitive benchmarks. Fourth, we primarily evaluated datasets involving within-dataset deduplication of demographic records. Performance on other kinds of data, such as citation networks, remains to be established.

## CONCLUSION

In summary, graph-based post-processing reduced linkage error rates by up to 66% relative to transitive closure, the standard approach. The top-performing algorithms – Leiden (modularity and CPM), Louvain, and label propagation – are readily available in existing software packages, computationally efficient, readily parallelizable, and scalable to large datasets. Graph-based post-processing can be applied after scoring candidate matches in any entity resolution pipeline. The improvements over transitive closure are large and come at little additional cost.

## Supplementary tables and figures

## Description of clustering algorithms

### Clustering algorithms previously systematically evaluated in entity resolution

**Transitive closure.** Transitive closure serves as the baseline against which all other algorithms are compared. Given a fixed threshold $\theta$, an edge $(r_i, r_j)$ is classified as a link (i.e., retained) if $s_{ij} \geq \theta$, and a non-link otherwise. The threshold $\theta$ is typically set to the value that minimizes false negative and false positive rates. Clusters are defined as the connected components of the resulting graph. Therefore, two records are assigned to the same individual if there exist a path of edges connecting them in the resulting graph. Transitive closure is deterministic and each run will yield the same partition.

**Center.** The center algorithm[1] partitions the similarity graph into clusters, with each cluster having a center node. Edges with $s_{ij} \geq \theta$ are first retained and then sorted in descending order of similarity score. The algorithm then performs a single scan of the sorted edge list. The first time a node $v_i$ is encountered, it is designated as a cluster center. All subsequent nodes $v_j$ that appear as a pair $(v_i, v_j)$ in the list are then assigned to the cluster of $v_i$. As a result, a node is only assigned to a cluster if has a direct edge with the center, thus resulting in more clusters than transitive closure. Center is deterministic and each run will yield the same partition.

**Merge-Center.** Merge-center[2] extends the center algorithm by introducing a merging step. When a node assigned to a cluster $C_i$ shares a direct edge with the center of another cluster $C_j$, the two clusters are merged, with the resulting cluster containing multiple center nodes. The merging step reduces the number of clusters relative to center. Merge-center is deterministic and each run will yield the same partition.

**Markov clustering (MCL).** The Markov clustering algorithm detects cluster structure by simulating random walks on the graph through iterative application of two operators: expansion and inflation[3]. The idea is that random walks on the graph will tend to stay within natural clusters, as there are many paths connecting nodes within a cluster and few paths leading out. Expansion raises the stochastic matrix to a power, computing longer random walks and increasing the probability of paths within clusters relative to paths between clusters. Inflation then rescales the resulting probabilities, strengthening intra-cluster flow while weakening inter-cluster flow. Iterating these two steps until convergence causes the separation of the graph into distinct components, each interpreted as a cluster. The granularity of the resulting clusters is controlled by the inflation parameter $r$, with larger values producing tighter clusters. Markov clustering is known to be very fast and scalable in large graphs. Markov clustering is deterministic and will yield the same partition across runs.

**Weighted diameter.** The weighted diameter algorithm is motivated by the heuristic that a cluster should be split if its two most dissimilar nodes are unlikely to refer to the same individual[4]. The algorithm initializes clusters as the connected components of the graph and iteratively removes the lowest-scoring edge until the weighted diameter of each cluster, i.e., the distance between its two most dissimilar nodes, falls below a chosen distance metric $d$, with higher values of $d$ indicating greater dissimilarity. As a result, every record pair within a resulting cluster is guaranteed to have a dissimilarity no greater than $d$, ensuring all records plausibly refer to the same individual.

**Clustering algorithms not yet systematically evaluated in entity resolution**

**Louvain (Multilevel).** The Louvain (i.e., multilevel) algorithm is a heuristic method for detecting cluster (i.e., community) structure that optimizes a global objective function, the modularity equation $Q$, through a series of local node assignments. It is widely regarded as intuitive, easy to implement, and extremely scalable[5]. Briefly, $Q$ measures how much the density of edges within clusters deviates from what would be expected under a degree-matched random graph[6]. The resolution parameter $\gamma > 0$ controls the granularity of resulting clusters, with higher values yielding more, smaller clusters.

Modularity $Q$ is defined as:

$$Q = \frac{1}{2m} \sum_{i,j} \left[ A_{ij} w_{ij} - \gamma \frac{k_i, k_j}{2m} \right] \delta(c_i, c_j) \,. \tag{1}$$

Where $A_{ij}$ is the adjacency matrix, with $A_{ij} = 1$ if an edge exists between nodes $v_i$ and $v_j$, and 0 otherwise, $w_{ij}$ is the weight of the edge between nodes $v_i$ and $v_j$, $k_i = \sum_j A_{ij}$ is the sum of the weight of edges attached to node $v_i$, $c_i$ is the cluster to which node $v_i$ is assigned, $\delta(c_i, c_j) = 1$ if $v_i$ and $v_j$ belong to the same cluster and $0$ otherwise, and $m = \frac{1}{2} \sum_{ij} A_{ij}$ is the total weight of all edges in the graph.

Louvain proceeds in two iteratively repeated phases. In the first phase (local moving of nodes), each node is initially assigned to its own cluster. For each node $v_i$, the algorithm evaluates the modularity gain $\Delta Q$ from moving $v_i$ to each neighboring cluster, placing $v_i$ in the cluster yielding the largest positive gain. This process is repeated until no individual move can improve $Q$. In the second phase (aggregation of the network), all nodes within the same cluster are aggregated into a single node, with edge weights between nodes in the aggregate network equal to the sum of weights between the nodes contained in each cluster in the original network. The two phases are then repeated until no further improvement in $Q$ is possible. Louvain is stochastic and may produce different partitions across runs.

Louvain has two notable limitations. First, modularity optimization is subject to resolution limit, where clusters smaller than a certain scale may be inappropriately merged.[7] Second, the aggregation step can produce internally disconnected clusters, i.e., nodes assigned to the same cluster despite no path connecting them within the cluster.[8] Both limitations motivate the development of Leiden algorithm, described below.

**Leiden.** The Leiden algorithm was developed to directly address the two key limitations of Louvain: 1) resolution limit of modularity; and 2) possibility of internally disconnected clusters[8]. Leiden optimizes either modularity or the Constant Potts Model (CPM) global objective function through local moving of nodes[8]. In CPM, the resolution parameter $\gamma$ has a straightforward interpretation: clusters should have an internal edge density of at least $\gamma$, with between-cluster density below $\gamma$. Higher values of $\gamma$ yields more, smaller clusters. Unlike modularity, CPM is resolution-limit-free -- clusters of any size can in principle be detected, as clustering decisions depend only on local edge density.

CPM is defined as:

$$\mathcal{H} = -\sum_{i,j} \left(A_{ij} w_{ij} - \gamma\right) \delta\left(c_i, c_j\right), \tag{2}$$

with $A_{ij}$, $w_{ij}$, and $\delta(c_i, c_j)$ as defined in (1).

Leiden extends Louvain by introducing a refinement phase, resulting in three phases: (1) local moving of nodes, (2) refinement of the partition, and (3) aggregation of the network. The local moving phase mirrors Louvain but uses a fast local move procedure that revisits only nodes whose neighborhood has changed. In the refinement phase, clusters identified in the local moving phase are further examined and potentially split into subcommunities, resulting in greater flexibility to find high-quality partitions. The aggregate network is then constructed based on the refined partition. Leiden guarantees that all resulting clusters are internally connected, and when run iteratively, converges to a partition in which all subsets of all clusters are locally optimally assigned. Leiden is stochastic and may produce different partitions across runs.

**Infomap.** Infomap approaches clustering as a compression problem – the optimal partition of a network is the one that minimizes the description length of a random walk on the graph, capturing how structure shapes the flow of information across the network [9,10] The idea is that a random walker tends to remain within densely connected clusters for extended periods before moving elsewhere, and a good partition exploits this regularity to describe the walk more concisely. This is quantified by the map equation, which measures the theoretical limit of how concisely a random walk can be described under a given partition. Infomap is stochastic and may yield different partitions across runs.

**Label propagation.** Label propagation is a localized, near-linear time algorithm that requires no predefined objective function[11]. Each node is first initialized with a unique label and iteratively adopts the label held by the majority of its neighbors (i.e., nodes that are directly connected by a single edge), with ties broken uniformly at random and nodes updated in random order. Densely connected groups of nodes quickly reach a consensus on a shared label, which then expands outward. The algorithm terminates when every node holds the most common label among its neighbors. Nodes sharing the same label are assigned to the same cluster. The algorithm may occasionally produce disconnected clusters when labels propagate in different directions. Label propagation is stochastic and may produce different partitions across runs.

**Fast greedy.** Fast greedy optimizes the global modularity objective function $Q$ in $(1)$ through a greedy agglomerative procedure[12]. Starting with each node in its own cluster, the algorithm repeatedly merges the two adjacent clusters whose amalgamation yields the greatest increase in $Q$. This process continues until no further improvement in $Q$ is possible, and the partition corresponding to the maximum $Q$ across all merge steps is returned as the final clustering. The full merging sequence can be represented as a dendrogram, capturing the hierarchical cluster structure at all levels. Fast greedy scales reasonably well to networks of approximately one million nodes. A known weakness is its tendency to produce overly large clusters containing a disproportionate fraction of nodes[5], even on networks with no significant cluster structure, which can degrade partition quality. Fast greedy is deterministic and yields the same partition across runs.

**Walktrap.** Walktrap detects cluster structure by using short random walks as a measure of structural similarity between nodes[13]. The intuition is that short random walks of length $t$ tend to remain within densely connected clusters, so two nodes that produce similar random walk probability distributions, i.e., that are likely to be visited by walks starting from similar locations, are likely in the same cluster. These distances are then used in a greedy agglomerative hierarchical clustering procedure: starting with each node in its own cluster, the algorithm iteratively merges the two clusters whose amalgamation minimizes the mean distance between nodes and their cluster, continuing until all nodes are in a single cluster. The partition corresponding to the maximum modularity $Q$ in $(1)$ across all steps of the dendrogram is returned as the final clustering. Walktrap is stochastic and may produce different partitions across runs.

**Notes for Tables S1-S5 A):** Merge-center has no tuning parameter but is typically applied after transitive closure; the reported optimal parameter for merge-center is the transitive closure threshold yielding the best performance after applying merge-center.

**Table S1. Validation performance of the FEBRL3 dataset, A) without adaptation to entity resolution settings; B) with adaptation to entity resolution settings**

**A)**

| Algorithm | Optimal parameter | PPV (%)* | Sen (%)* | F-score (%)* | Runtime (ms) |
|---|---|---|---|---|---|
| Gold standard | - | 100.000 (100.000, 100.000) | 98.753 (98.547, 98.897) | 99.373 (99.268, 99.445) | 38 |
| Transitive closure | 0.3 ($\theta$) | 99.817 (99.769, 99.831) | 97.327 (97.073, 97.590) | 98.556 (98.433, 98.694) | 16 |
| Center | - | 99.258 (99.078, 99.400) | 90.847 (90.557, 91.030) | 94.866 (94.768, 94.983) | 28 |
| Merge-Center | 0.2 ($\theta$) | 99.707 (99.635, 99.727) | 96.620 (96.500, 96.817) | 98.139 (98.087, 98.247) | 41 |
| Markov | 1.6 ($r$) | 99.764 (99.700, 99.831) | 98.047 (97.787, 98.227) | 98.898 (98.754, 98.990) | 3,495 |
| Weighted diameter | 0.2 ($\theta$) | 99.426 (99.335, 99.490) | 97.727 (97.517, 97.863) | 98.569 (98.462, 98.670) | 937 |
| Fastgreedy | - | 96.107 (95.816, 96.485) | 98.753 (98.547, 98.897) | 97.412 (97.214, 97.674) | 43 |
| Infomap | - | 98.190 (97.957, 98.443) | 98.753 (98.547, 98.897) | 98.471 (98.307, 98.670) | 410 |
| Label propagation | - | 98.773 (98.606, 98.933) | 98.753 (98.547, 98.897) | 98.763 (98.636, 98.915) | 21 |
| Leiden (CPM) | 0.11 ($\gamma$) | 99.818 (99.734, 99.865) | 98.080 (97.820, 98.260) | 98.941 (98.787, 99.056) | 25 |
| Leiden (modularity) | 150 ($\gamma$) | 98.641 (98.474, 98.800) | 98.753 (98.547, 98.897) | 98.697 (98.570, 98.849) | 24 |
| Louvain | 150 ($\gamma$) | 98.641 (98.474, 98.800) | 98.753 (98.547, 98.897) | 98.697 (98.570, 98.849) | 19 |
| Walktrap | 1 ($t$) | 96.107 (95.816, 96.485) | 98.753 (98.547, 98.880) | 97.412 (97.214, 97.674) | 228 |

*Mean (2.5th-97.5th percentile) across 5 runs with random reference nodes*

**B)**

| Algorithm | Optimal parameter | $\theta_{low}$ | $\theta_{high}$ | PPV (%)* | Sen (%)* | F-score (%)* | Runtime (ms) |
|---|---|---|---|---|---|---|---|
| Gold standard | - | - | - | 100.000 (100.000, 100.000) | 98.753 (98.547, 98.897) | 99.373 (99.268, 99.445) | 38 |
| Transitive closure | 0.3 ($\theta$) | - | - | 99.817 (99.769, 99.831) | 97.327 (97.073, 97.590) | 98.556 (98.433, 98.694) | 16 |
| Center | - | 0.0 | 0.4 | 99.269 (99.109, 99.420) | 98.547 (98.290, 98.700) | 98.907 (98.756, 99.040) | 47 |
| Merge-Center | - | 0.2 | 0.4 | 99.709 (99.638, 99.729) | 97.640 (97.440, 97.797) | 98.664 (98.571, 98.750) | 41 |
| Markov | 1.6 ($r$) | 0.1 | 0.4 | 99.818 (99.733, 99.865) | 98.047 (97.787, 98.227) | 98.924 (98.770, 99.040) | 3,462 |
| Weighted diameter | 0.2 ($\theta$) | - | - | 99.426 (99.335, 99.490) | 97.727 (97.517, 97.863) | 98.569 (98.462, 98.670) | 937 |
| Fastgreedy | - | 0.3 | 0.0** | 99.817 (99.769, 99.831) | 97.327 (97.073, 97.590) | 98.556 (98.433, 98.694) | 47 |
| Infomap | - | 0.2 | 0.9 | 99.682 (99.598, 99.730) | 97.727 (97.517, 97.863) | 98.695 (98.611, 98.781) | 409 |
| Label propagation | - | 0.1 | 0.4 | 99.428 (99.270, 99.529) | 98.140 (97.860, 98.300) | 98.780 (98.648, 98.903) | 28 |
| Leiden (CPM) | 0.11 ($\gamma$) | 0.0 | 0.4 | 99.818 (99.733, 99.865) | 98.080 (97.820, 98.260) | 98.941 (98.787, 99.056) | 35 |
| Leiden (modularity) | 150 ($\gamma$) | 0.1 | 0.3 | 99.294 (99.136, 99.395) | 98.140 (97.860, 98.300) | 98.714 (98.582, 98.836) | 36 |
| Louvain | 150 ($\gamma$) | 0.1 | 0.3 | 99.294 (99.136, 99.395) | 98.140 (97.860, 98.300) | 98.714 (98.582, 98.836) | 33 |
| Walktrap | 1 ($t$) | 0.3 | 0.0** | 99.817 (99.769, 99.831) | 97.327 (97.073, 97.590) | 98.556 (98.433, 98.694) | 121 |

*Mean (2.5th-97.5th percentile) across 5 runs with random reference nodes

** These are no different from doing a transitive closure

**Table S2. Validation performance of the Historical 50K dataset, A) without adaptation to entity resolution settings; B) with adaptation to entity resolution settings**

**A)**

| Algorithm | Optimal parameter | PPV (%)* | Sen (%)* | F-score (%)* | Runtime (ms) |
|---|---|---|---|---|---|
| Gold standard | - | 100.000 (100.000, 100.000) | 92.457 (92.117, 92.652) | 96.081 (95.897, 96.186) | 399 |
| Transitive closure | 0.93 ($\theta$) | 94.265 (94.163, 94.353) | 81.530 (81.131, 81.792) | 87.436 (87.212, 87.605) | 231 |
| Center | - | 99.095 (99.044, 99.168) | 56.188 (55.717, 56.699) | 71.713 (71.327, 72.117) | 903 |
| Merge-Center | 0.7 ($\theta$) | 90.730 (90.622, 90.795) | 81.832 (81.454, 82.215) | 86.051 (85.856, 86.271) | 577 |
| Markov | 2.0 ($r$) | 96.591 (96.440, 96.805) | 88.108 (87.793, 88.333) | 92.155 (91.960, 92.318) | 29,716 |
| Weighted diameter | 0.82 ($\theta$) | 92.909 (92.749, 93.014) | 83.493 (83.208, 83.734) | 87.949 (87.770, 88.099) | 14,573 |
| Fastgreedy | - | 54.789 (54.655, 54.945) | 92.514 (92.171, 92.706) | 68.821 (68.685, 68.986) | 2,337 |
| Infomap | - | 89.462 (89.193, 89.773) | 91.990 (91.671, 92.190) | 90.708 (90.526, 90.951) | 8,879 |
| Label propagation | - | 94.768 (94.565, 95.025) | 91.331 (90.994, 91.600) | 93.018 (92.765, 93.281) | 1,064 |
| Leiden (CPM) | 0.06 ($\gamma$) | 92.683 (92.459, 92.914) | 89.061 (88.741, 89.324) | 90.836 (90.606, 91.065) | 505 |
| Leiden (modularity) | 800 ($\gamma$) | 94.920 (94.652, 95.187) | 91.373 (90.934, 91.617) | 93.112 (92.837, 93.368) | 506 |
| Louvain | 800 ($\gamma$) | 94.904 (94.662, 95.196) | 91.402 (91.020, 91.633) | 93.120 (92.866, 93.381) | 553 |
| Walktrap | 3 ($t$) | 81.455 (81.153, 81.747) | 91.195 (90.810, 91.408) | 86.050 (85.839, 86.295) | 19,050 |

**Mean (2.5th-97.5th percentile) across 5 runs with random reference nodes*

**B)**

| Algorithm | Optimal parameter | $\theta_{low}$ | $\theta_{high}$ | PPV (%)* | Sen (%)* | F-score (%)* | Runtime (ms) |
|---|---|---|---|---|---|---|---|
| Gold standard | - | - | - | 100.000 (100.000, 100.000) | 92.457 (92.117, 92.652) | 96.081 (95.897, 96.186) | 399 |
| Transitive closure | 0.93 ($\theta$) | - | - | 94.265 (94.163, 94.353) | 81.530 (81.131, 81.792) | 87.436 (87.212, 87.605) | 231 |
| Center | - | 0.0 | 0.99 | 97.039 (96.987, 97.093) | 86.773 (86.383, 87.023) | 91.619 (91.391, 91.766) | 893 |
| Merge-Center** | - | 0.9 | 0.9 | 93.012 (92.929, 93.076) | 82.479 (82.149, 82.740) | 87.430 (87.229, 87.602) | 773 |
| Markov | 2.0 ($r$) | 0.0 | 1.0 | 96.591 (96.440, 96.805) | 88.108 (87.793, 88.333) | 92.155 (91.960, 92.318) | 30,141 |
| Weighted diameter | 0.82 ($\theta$) | - | - | 92.909 (92.749, 93.014) | 83.493 (83.208, 83.734) | 87.949 (87.770, 88.099) | 14,573 |
| Fastgreedy | - | 0.9 | 0.99 | 93.012 (92.929, 93.076) | 82.479 (82.149, 82.740) | 87.430 (87.229, 87.602) | 2,745 |
| Infomap | - | 0.0 | 0.9995 | 89.437 (89.232, 89.741) | 91.974 (91.667, 92.154) | 90.687 (90.548, 90.914) | 9,214 |
| Label propagation | - | 0.0 | 1.0 | 94.620 (94.421, 94.850) | 91.373 (91.075, 91.613) | 92.968 (92.718, 93.203) | 1,112 |
| Leiden (CPM) | 0.1 ($\gamma$) | 0.0 | 0.995 | 94.341 (94.136, 94.542) | 87.614 (87.219, 87.823) | 90.853 (90.547, 90.969) | 794 |
| Leiden (modularity) | 1200 ($\gamma$) | 0.0 | 0.9995 | 95.176 (95.021, 95.308) | 91.173 (90.794, 91.433) | 93.131 (92.930, 93.330) | 746 |
| Louvain | 1200 ($\gamma$) | 0.0 | 0.9995 | 95.077 (94.904, 95.244) | 91.201 (90.812, 91.437) | 93.099 (92.878, 93.301) | 726 |
| Walktrap | 3 ($t$) | 0.8 | 0.99 | 89.516 (89.381, 89.613) | 84.619 (84.331, 84.833) | 86.999 (86.866, 87.153) | 16,392 |

**Mean (2.5th-97.5th percentile) across 5 runs with random reference nodes*

***Note: in the Merge-Center case, $\theta_{low} = \theta_{high}$. This is equivalent to doing transitive closure (i.e., Merge-center led to no improvement over transitive closure)*

**Table S3. Validation performance of the Splink 7m dataset, A) without adaptation to entity resolution settings; B) with adaptation to entity resolution settings**

**A)**

| Algorithm | Optimal parameter | PPV (%)* | Sen (%)* | F-score (%)* | Runtime (ms) |
|---|---|---|---|---|---|
| Gold standard | - | 100.000 (100.000, 100.000) | 94.796 (94.772, 94.823) | 97.329 (97.316, 97.342) | 81,428 |
| Transitive closure | 0.97 ($\theta$) | 96.989 (96.977, 97.007) | 90.831 (90.815, 90.857) | 93.809 (93.800, 93.819) | 44,195 |
| Center | - | 99.337 (99.330, 99.347) | 48.551 (48.531, 48.581) | 65.224 (65.208, 65.250) | 124,684 |
| Merge-Center | 0.9 ($\theta$) | 95.909 (96.898, 95.925) | 90.810 (90.789, 90.832) | 93.290 (93.277, 93.303) | 78,634 |
| Markov | - | | | | - |
| Weighted diameter | 0.94 ($\theta$) | 96.660 (96.650, 96.672) | 91.568 (91.552, 91.592) | 94.045 (94.036, 94.056) | 2,549,882 |
| Fastgreedy | - | 50.581 (50.574, 50.592) | 94.923 (94.902, 94.948) | 65.996 (65.985, 66.004) | 53,331,589 |
| Infomap | - | 90.164 (90.145, 90.185) | 94.557 (94.533, 94.582) | 92.308 (92.289, 92.329) | 1,617,669 |
| Label propagation | - | 96.667 (96.642, 96.689) | 93.797 (93.777, 93.815) | 95.210 (95.195, 95.227) | 524,501 |
| Leiden (CPM) | 0.12 ($\gamma$) | 97.328 (97.315, 97.342) | 92.417 (92.403, 92.432) | 94.809 (94.796, 94.818) | 131,047 |
| Leiden (modularity) | 150,000 ($\gamma$) | 96.793 (96.778, 96.811) | 93.598 (93.585, 93.609) | 95.168 (95.155, 95.182) | 117,871 |
| Louvain | 150,000 ($\gamma$) | 96.761 (96.748, 96.779) | 93.615 (93.599, 93.629) | 95.162 (95.148, 95.177) | 130,023 |
| Walktrap | - | | | | - |

*Mean (2.5$^{th}$-97.5$^{th}$ percentile) across 5 runs with random reference nodes*

**B)**

| Algorithm | Optimal parameter | $\theta_{low}$ | $\theta_{high}$ | PPV (%)* | Sen (%)* | F-score (%)* | Runtime (ms) |
|---|---|---|---|---|---|---|---|
| Gold standard | - | - | - | 100.000 (100.000, 100.000) | 94.796 (94.772, 94.823) | 97.329 (97.316, 97.342) | 81,428 |
| Transitive closure | 0.97 ($\theta$) | - | - | 96.989 (96.977, 97.007) | 90.831 (90.815, 90.857) | 93.809 (93.800, 93.819) | 44,195 |
| Center | - | 0.0 | 0.99 | 97.373 (97.362, 97.386) | 92.214 (92.195, 92.231) | 94.724 (94.712, 94.733) | 147,156 |
| Merge-Center | - | 0.9 | 0.99 | 95.538 (95.528, 95.554) | 92.018 (92.007, 92.038) | 93.745 (93.736, 93.752) | 117,940 |
| Markov | - | - | - | | | | - |
| Weighted diameter | 0.94 ($\theta$) | - | - | 96.660 (96.650, 96.672) | 91.568 (91.552, 91.592) | 94.045 (94.036, 94.056) | 2,549,882 |
| Fastgreedy | - | 0.9 | 0.999 | 94.049 (94.038, 94.069) | 92.452 (92.441, 92.471) | 93.244 (93.236, 93.254) | 86,101,329 |
| Infomap | - | 0.9 | 1.0 | 97.257 (97.244, 97.272) | 92.384 (92.372, 92.402) | 94.758 (94.747, 94.769) | 1,358,661 |
| Label propagation | - | 0.7 | 1.0 | 98.383 (98.370, 98.392) | 92.794 (92.774, 92.819) | 95.507 (95.496, 95.521) | 359,263 |
| Leiden (CPM) | 0.12 ($\gamma$) | 0.7 | 0.9995 | 97.745 (97.733, 97.756) | 92.681 (92.663, 92.697) | 95.146 (95.131, 95.158) | 143,394 |
| Leiden (modularity) | 150,000 ($\gamma$) | 0.7 | 0.9999 | 97.932 (97.921, 97.942) | 93.308 (93.288, 93.334) | 95.564 (95.548, 95.582) | 138,850 |
| Louvain | 150,000 ($\gamma$) | 0.7 | 0.9999 | 97.909 (97.898, 97.919) | 93.309 (93.289, 93.337) | 95.554 (95.538, 95.572) | 148,628 |
| Walktrap | - | - | - | - | - | - | - |

**Mean (2.5th-97.5th percentile) across 5 runs with random reference nodes*

**Table S4. Validation performance of the NHLS HIV cohort dataset, A) without adaptation to entity resolution settings; B) with adaptation to entity resolution settings**

**A)**

| Algorithm | Optimal parameter | PPV (%)* | Sen (%)* | F-score (%)* | Runtime (ms) |
|---|---|---|---|---|---|
| Gold standard | - | 99.935 (99.933, 99.936) | 96.873 (96.844, 96.905) | 98.380 (98.365, 98.397) | 21,579 |
| Transitive closure | 0.6 ($\theta$) | 95.086 (95.065, 95.110) | 92.663 (92.599, 92.703) | 93.859 (93.816, 93.889) | 18,413 |
| Center | - | 96.961 (96.937, 96.988) | 77.643 (77.613, 77.680) | 86.233 (86.213, 86.262) | 63,529 |
| Merge-Center | 0.5 ($\theta$) | 93.394 (93.383, 93.404) | 92.290 (92.234, 92.348) | 92.839 (92.805, 92.868) | 74,844 |
| Markov | - | | | | - |
| Weighted diameter | 0.5 ($\theta$) | 96.022 (95.981, 96.063) | 93.423 (93.380, 93.475) | 94.704 (94.664, 94.750) | 3,817,950 |
| Fastgreedy | - | 60.893 (60.830, 60.969) | 99.356 (99.343, 99.366) | 75.509 (75.460, 75.567) | 61,548,062 |
| Infomap | - | 80.846 (80.816, 80.870) | 98.995 (98.987, 99.001) | 89.005 (88.988, 89.022) | 3,661,127 |
| Label propagation | - | 93.004 (92.970, 93.032) | 95.404 (95.359, 95.435) | 94.189 (94.150, 94.203) | 630,679 |
| Leiden (CPM) | 0.12 ($\gamma$) | 95.017 (94.989, 95.040) | 94.994 (94.975, 95.015) | 95.005 (94.998, 95.019) | 135,029 |
| Leiden (modularity) | 200,000 ($\gamma$) | 89.268 (89.251, 89.300) | 95.979 (95.962, 96.007) | 92.502 (92.488, 92.516) | 103,730 |
| Louvain | 200,000 ($\gamma$) | 89.123 (89.089, 89.149) | 96.032 (95.991, 96.074) | 92.449 (92.421, 92.466) | 102,891 |
| Walktrap | - | - | - | - | - |

*Mean (2.5th-97.5th percentile) across 5 runs with random reference nodes*

**B)**

| Algorithm | Optimal parameter | $\theta_{low}$ | $\theta_{high}$ | PPV (%)* | Sen (%)* | F-score (%)* | Runtime (ms) |
|---|---|---|---|---|---|---|---|
| Gold standard | - | - | - | 99.935 (99.933, 99.936) | 96.873 (96.844, 96.905) | 98.380 (98.365, 98.397) | 21,579 |
| Transitive closure | 0.6 ($\theta$) | - | - | 95.086 (95.065, 95.110) | 92.663 (92.599, 92.703) | 93.859 (93.816, 93.889) | 18,413 |
| Center | - | 0.4 | 0.7 | 95.628 (95.580, 95.654) | 93.532 (93.516, 93.551) | 94.568 (94.554, 94.580) | 90,051 |
| Merge-Center | - | 0.6 | 0.7 | 95.676 (95.651, 95.707) | 92.274 (92.228, 92.302) | 93.944 (93.918, 93.964) | 118,507 |
| Markov | - | - | - | | | | - |
| Weighted diameter | 0.5 ($\theta$) | - | - | 96.022 (95.981, 96.063) | 93.423 (93.380, 93.475) | 94.704 (94.664, 94.750) | 3,817,950 |
| Fastgreedy | - | 0.6 | 1.01 | 95.088 (95.046, 95.125) | 92.673 (92.651, 92.713) | 93.865 (93.835, 93.893) | 95,424,246 |
| Infomap | - | 0.6 | 0.99 | 96.342 (96.321, 96.354) | 92.561 (92.528, 92.589) | 94.414 (94.402, 94.427) | 2,549,721 |
| Label propagation | - | 0.3 | 0.99 | 95.833 (95.816, 95.847) | 94.511 (94.489, 94.553) | 95.167 (95.157, 95.187) | 568,463 |
| Leiden (CPM) | 0.08 ($\gamma$) | 0.4 | 0.99 | 95.933 (95.896, 95.959) | 95.115 (95.081, 95.156) | 95.522 (95.501, 95.555) | 126,557 |
| Leiden (modularity) | 180,000 ($\gamma$) | 0.4 | 0.99 | 95.035 (95.011, 95.056) | 95.418 (95.397, 95.447) | 95.226 (95.213, 95.249) | 122,630 |
| Louvain | 180,000 ($\gamma$) | 0.4 | 0.99 | 94.966 (94.902, 94.999) | 95.472 (95.430, 95.503) | 95.218 (95.167, 95.237) | 170,824 |
| Walktrap | - | - | - | - | - | - | - |

*Mean (2.5th-97.5th percentile) across 5 runs with random reference nodes*

**Table S5. Validation performance of the NHLS HIV cohort dataset (test set), with adaptation to entity resolution settings**

| Algorithm | Optimal parameter | $\theta_{low}$ | $\theta_{high}$ | PPV (%) | Sen (%) | F-score (%) | Runtime (ms) |
|---|---|---|---|---|---|---|---|
| Gold standard | - | - | - | 99.931 (99.929, 99.932) | 96.938 (96.904, 96.962) | 98.411 (98.393, 98.424) | 29,452 |
| Transitive closure | 0.6 ($\theta$) | - | - | 95.100 (95.082, 95.130) | 92.714 (92.641, 92.769) | 93.892 (93.850, 93.923) | 22,830 |
| Center | - | 0.4 | 0.7 | 95.546 (95.526, 95.564) | 93.603 (93.574, 93.621) | 94.564 (94.544, 94.582) | 134,587 |
| Merge-Center | - | 0.6 | 0.7 | 95.612 (95.577, 95.649) | 92.294 (92.281, 92.312) | 93.924 (93.900, 93.944) | 116,130 |
| Markov | - | - | - | | | | - |
| Weighted diameter | 0.5 | - | - | 96.031 (96.006, 96.065) | 93.442 (93.398, 93.480) | 94.719 (94.687, 94.735) | 3,803,670 |
| Fastgreedy | - | - | - | | | | - |
| Infomap | - | - | - | 96.328 (96.280, 96.375) | 92.658 (92.634, 92.696) | 94.458 (94.424, 94.489) | 251688175 |
| Label propagation | - | 0.3 | 0.99 | 95.751 (95.713, 95.791) | 94.533 (94.515, 94.548) | 95.138 (95.121, 95.159) | 943,050 |
| Leiden (CPM) | 0.08 ($\gamma$) | 0.4 | 0.99 | 95.902 (95.885, 95.921) | 95.146 (95.112, 95.170) | 95.523 (95.501, 95.538) | 240,686 |
| Leiden (modularity) | 180,000 ($\gamma$) | 0.4 | 0.99 | 95.081 (95.065, 95.111) | 95.449 (95.384, 95.501) | 95.265 (95.229, 95.299) | 300,891 |
| Louvain | 180,000 ($\gamma$) | 0.4 | 0.99 | 95.041 (95.020, 95.057) | 95.477 (95.432, 95.510) | 95.258 (95.231, 95.278) | 222,826 |
| Walktrap | - | - | - | - | - | - | - |

**Table S6. Algorithms considered and inclusion/exclusion criteria**

| Algorithm | Graph as input | Partition into disjoint clusters | Code availability | Scalable | No inestimable input parameters | Meets criteria |
|---|---|---|---|---|---|---|
| Transitive closure | Yes | Yes | Pseudocode | Yes | Yes | Yes |
| Leiden[8] | Yes | Yes | iGraph package | Yes | Yes | Yes |
| Louvain[5] | Yes | Yes | iGraph package | Yes | Yes | Yes |
| Walktrap[13] | Yes | Yes | iGraph package | Yes | Yes | Yes |
| Infomap[9,10] | Yes | Yes | iGraph package | Yes | Yes | Yes |
| Label propagation[11] | Yes | Yes | iGraph package | Yes | Yes | Yes |
| Fast greedy[12] | Yes | Yes | iGraph package | Yes | Yes | Yes |
| Center[14] | Yes | Yes | Pseudocode | Yes | Yes | Yes |
| Merge-center[2] | Yes | Yes | Pseudocode | Yes | Yes | Yes |
| Markov[15] | Yes | Yes | Markov_ clustering package | Yes | Yes | Yes |
| Weighted diameter[4] | Yes | Yes | Code from authors | Yes | Yes | Yes |
| Leading eigenvector*[16] | Yes | Yes | iGraph package | No | Yes | No |
| Star[17] | Yes | No | Pseudocode | Yes | Yes | No |
| Articulation point[18] | Yes | No | No accessible code | Yes | Yes | No |
| Optimal modularity[19] | Yes | Yes | iGraph package | No | Yes | No |
| Ricochet[20] | Yes | Yes (some versions) | No accessible code | No | Yes | No |
| Correlation clustering[21] | Yes | Yes | No accessible code | No | Yes | No |
| Cut clustering[22] | Yes | Yes | No accessible code | No | No | No |
| Extended Maximum Clique Clustering[23] | Yes | Yes | No accessible code | No | Yes | No |
| G-Cluster[24] | Yes | Yes | No accessible code | No | Yes | No |
| H-Cluster[24] | Yes | Yes | No accessible code | Yes | Yes | No |
| Spinglass[25,26] | Yes | Yes | iGraph package | No | Yes | No |
| Edge betweenness[27] | Yes | Yes | iGraph package | No | Yes | No |
| K-means[28] | No | Yes | Scikit-learn package | Yes | No | No |
| DBSCAN[29] | No | No | Scikit-learn package | Yes | Yes | No |
| Gaussian Mixture Models[30] | No | No | Scikit-learn package | Yes | No | No |
| Spectral clustering[31] | No | Yes | Scikit-learn package | No | No | No |

**Excluded because its eigenvector computation frequently fails to converge on small, sparse components, which are the dominant graph structures in entity resolution.*

**Table S7. Stability of stochastic clusters in the NHLS HIV cohort, at best tuning parameters**

| Clustering approach | N (%) of stable clusters | N (%) of unstable clusters |
|---|---|---|
| Leiden (CPM) | 2,021,063 (99.3%) | 13,878 (0.7%) |
| Leiden (modularity) | 1,947,322 (99.5%) | 10,551 (0.5%) |
| Louvain | 1,930,317 (98.7%) | 25,387 (1.3%) |
| Label propagation | 1,829,446 (95.3%) | 90,042 (4.7%) |

**Figure S1. Structural properties of entity resolution settings that motivates adaptation.**

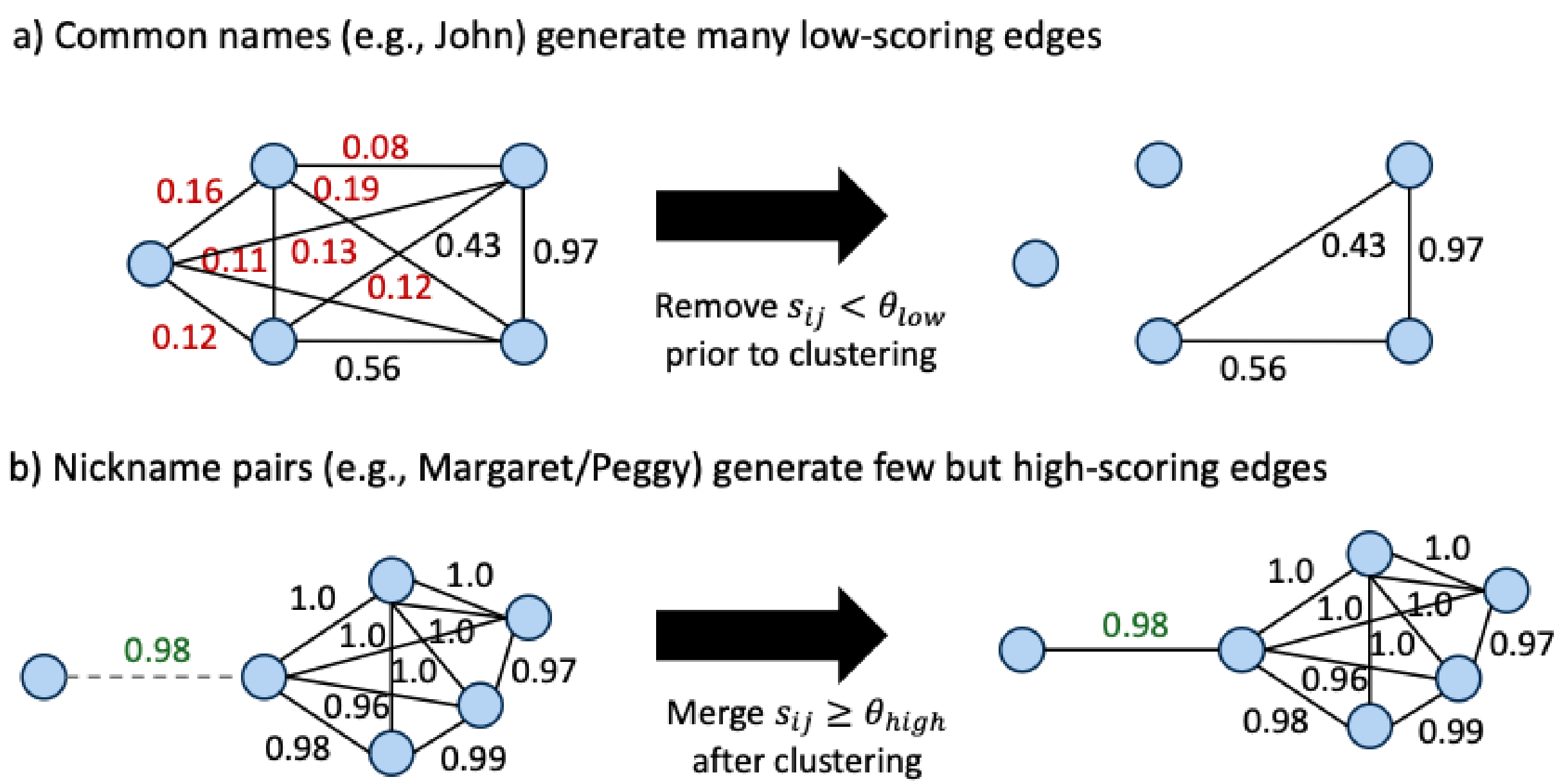

**Figure S2.** Illustration of how top-performing algorithms assign cluster membership.

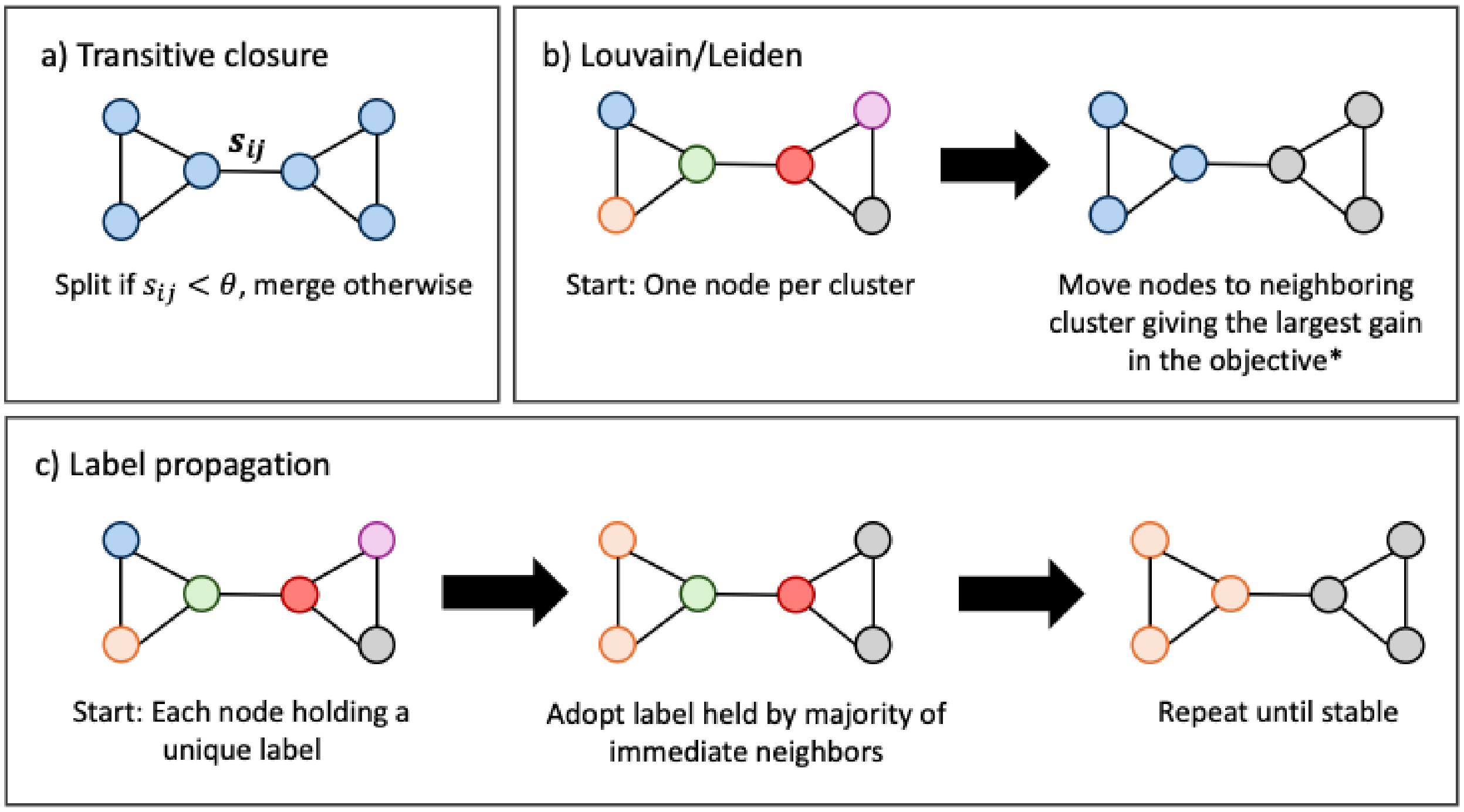

**Modularity: More internal edges than expected in a degree-matched random graph; CPM: internal density $> \gamma$, external density $< \gamma$*